\documentclass[12pt]{article}
\usepackage[left=2.5cm,top=2.50cm,right=2.5cm,bottom=2.50cm]{geometry}
\usepackage{mathrsfs}
\usepackage{amsmath,amssymb,latexsym,color,cancel,graphicx,bbm,colortbl}
\usepackage[english]{babel}
\usepackage[latin1]{inputenc}
\usepackage{ragged2e}
\begin{document}
\date{}

\title{The $SU(1,1)$ Perelomov number coherent states and the non-degenerate parametric amplifier}
\author{D. Ojeda-Guill\'en$^{a}$,\footnote{{\it E-mail address:} dojedag@ipn.mx}\\ R. D. Mota$^{b}$ and V. D. Granados$^{a}$} \maketitle

\begin{minipage}{0.9\textwidth}
\small $^{a}$ Escuela Superior de F{\'i}sica y Matem\'aticas,
Instituto Polit\'ecnico Nacional, Ed. 9, Unidad Profesional Adolfo L\'opez Mateos, C.P. 07738 M\'exico D. F., Mexico.\\

\small $^{b}$ Escuela Superior de Ingenier{\'i}a Mec\'anica y El\'ectrica, Unidad Culhuac\'an,
Instituto Polit\'ecnico Nacional, Av. Santa Ana No. 1000, Col. San Francisco Culhuac\'an, Delegaci\'on Coyoac\'an, C.P. 04430,  M\'exico D. F., Mexico.\\

\end{minipage}

\begin{abstract}
 We construct the Perelomov number coherent states for any three $su(1,1)$ Lie algebra generators
 and study some of their properties. We introduce three operators which act on Perelomov number
 coherent states and close the $su(1,1)$ Lie algebra. We show that the most general $SU(1,1)$
 coherence-preserving Hamiltonian has the Perelomov number coherent states as eigenfunctions,
 and we obtain their time evolution. We apply our results to obtain the non-degenerate parametric
 amplifier eigenfunctions, which are shown to be the Perelomov number coherent states of the
 two-dimensional harmonic oscillator.

\end{abstract}


\section{Introduction}
The harmonic oscillator coherent states can be constructed in three equivalent ways \cite{Nieto}: a) as
eigenstates of the annihilation operator, b) as the action of the displacement operator on the ground
state, or c) as the Heisenberg minimum uncertainty states.

The Klauder's displacement operator method \cite{Klau} was used to introduce
the displaced number states of the harmonic oscillator. Boiteux and Levelut defined these states as
the displacement operator applied to any state $|n\rangle$ of the harmonic oscillator and they called them
semicoherent states \cite{BandL}. After the introduction of the number coherent states, Roy and Singh \cite{Roy},
Satyanarayana \cite{Saty}, and Oliveira, Kim, Night and Bu$\check{z}$ek \cite{Oli} gave a detailed study of
the properties of these states. Nieto \cite{Nieto2} reviewed these properties and derived their most general form.
These coherent states are related to the Heisenberg-Weyl group.

Perelomov generalized the Klauder's approach for arbitrary Lie groups by defining
the coherent states as the action of the displacement operator on the ground states \cite{Perel}.
In this formulation the displacement operator depends on the Lie algebra generators. In general, coherent
states have been applied successfully to many physics fields, as can be seen in references \cite{Perellibro,Zhang,Klauderlibro,Gazeaulibro}.

Gerry applied the Perelomov displacement operator to any excited group states to study the Berry's phase in the degenerate parametric amplifier \cite{gerryberry}. In the optical parametric amplifier, one photon of a pump field transforms, via the nonlinear medium, into two photons called signal and idler. These output beams have the same frequency and polarization in the degenerate case and different ones in the non-degenerate case. This pairwise production of photons results in the conservation of the photon-number difference between the signal and idler modes in the absence of any loss. The high correlation between the signal and idler fields is responsible for the generation of a squeezed-vacuum state in the output of the device \cite{holmes}.

The aim of this work is to give some results on the theory of the $SU(1,1)$ Perelomov number coherent states. Given
any three $su(1,1)$ Lie algebra generators we obtain the Perelomov number coherent states. As an application of our results,
we obtain the eigenstates of the non-degenerate parametric amplifier \cite{louisell,mollow,mandel}. We show that these eigenstates
are the Perelomov number coherent states of the two-dimensional harmonic oscillator. To our knowledge these states have not been reported
previously.

This work is organized as it follows. In Section $2$ we give a summary on the unitary irreducible representation of the
$su(1,1)$ Lie algebra and the standard Perelomov coherent states (generated from the ground state).
Based on the works of Gerry \cite{gerryberry} and Huang \cite{huang}, we show that the most general $SU(1,1)$
coherence-preserving Hamiltonian has the Perelomov number coherent states as eigenfunctions. We construct
the Perelomov number coherent states for any three $su(1,1)$ Lie algebra generators. By applying similarity
transformations to the $su(1,1)$ Lie algebra generators by means of the displacement operator, we obtain
three new operators which act on the Perelomov number coherent states. We show that these operators
close the $su(1,1)$ Lie algebra and that the basis for its unitary irreducible representation is
given by the Perelomov number coherent states. We obtain the time evolution for the coherence-preserving
Hamiltonian eigenfunctions. In Section $3$, we apply our results to obtain the eigenfunctions of the non-degenerate
parametric amplifier, which result to be the Perelomov number coherent states of the
two-dimensional harmonic oscillator. Finally, we give some concluding remarks.

\section{$SU(1,1)$ generalized Perelomov coherent states}

\subsection{Standard $SU(1,1)$ Perelomov coherent states}

The $su(1,1)$ Lie algebra is spanned by the generators $K_{+}$, $K_{-}$
and $K_{0}$, which satisfy the commutation relations \cite{Vourdas}
\begin{eqnarray}
[K_{0},K_{\pm}]=\pm K_{\pm},\quad\quad [K_{-},K_{+}]=2K_{0}.\label{com}
\end{eqnarray}
The action of these operators on the Fock space states
$\{|k,n\rangle, n=0,1,2,...\}$ is
\begin{equation}
K_{+}|k,n\rangle=\sqrt{(n+1)(2k+n)}|k,n+1\rangle,\label{k+n}
\end{equation}
\begin{equation}
K_{-}|k,n\rangle=\sqrt{n(2k+n-1)}|k,n-1\rangle,\label{k-n}
\end{equation}
\begin{equation}
K_{0}|k,n\rangle=(k+n)|k,n\rangle,\label{k0n}
\end{equation}
where $|k,0\rangle$ is the lowest normalized state. The Casimir
operator $K^{2}=K^2_0-\frac{1}{2}(K_+K_-+K_-K_+)$ for any irreducible
representation satisfies $K^{2}=k(k-1)$. Thus, a representation of
$su(1,1)$ algebra is determined by the number $k$, called the Bargmann index. For the purpose of the
present work we will restrict to the discrete series only, for which
$k>0$.

The displacement operator $D(\xi)$ for this group is defined as
\begin{equation}
D(\xi)=\exp(\xi K_{+}-\xi^{*}K_{-}),
\end{equation}
where $\xi=-\frac{1}{2}\tau e^{-i\varphi}$, $-\infty<\tau<\infty$ and $0\leq\varphi\leq2\pi$. From
the properties $K^{\dag}_{+}=K_{-}$ and $K^{\dag}_{-}=K_{+}$ it can be shown that
the displacement operator possesses the property
\begin{equation}
D^{\dag}(\xi)=\exp(\xi^{*} K_{-}-\xi K_{+})=D(-\xi),
\end{equation}
and the so-called normal form of the displacement operator is given by
\begin{equation}
D(\xi)=\exp(\zeta K_{+})\exp(\eta K_{0})\exp(-\zeta^*K_{-})\label{normal},
\end{equation}
where  $\zeta=-\tanh(\frac{1}{2}\tau)e^{-i\varphi}$ and $\eta=-2\ln \cosh|\xi|=\ln(1-|\zeta|^2)$ \cite{Gerry}.
By using this normal form of the displacement operator and equations (\ref{k+n})-(\ref{k0n}), the standard $SU(1,1)$ coherent states are found to
be \cite{Perellibro}
\begin{equation}
|\zeta\rangle=D(\xi)|k,0\rangle=(1-|\zeta|^2)^k\sum_{s=0}^\infty\sqrt{\frac{\Gamma(n+2k)}{s!\Gamma(2k)}}\zeta^s|k,s\rangle.\label{PCN}
\end{equation}

\subsection{Motivation of the $SU(1,1)$ Perelomov number coherent states}

As a motivation for introducing the $SU(1,1)$ Perelomov number coherent states we consider the Hamiltonian
\begin{equation}
H=f K_0+\gamma \left(K_+e^{-i\phi}+K_-e^{i\phi}\right).\label{ham}
\end{equation}
This Hamiltonian is used to describe the degenerate and non-degenerate parametric amplifiers of quantum optics, the weakly
interacting Bose system, among others. Moreover, Gerry showed that this is the most general $SU(1,1)$ coherence-preserving
Hamiltonian \cite{Gerry}. The results in this subsection are a generalized form of those developed by Gerry \cite{gerryberry} and
Huang \cite{huang}.

In order to transform this Hamiltonian to a diagonal one in the basis of irreducible representation of the $su(1,1)$ Lie algebra, we applied the tilting transformation to the Schrödinger equation $H\Psi=E\Psi$ as it follows
\begin{equation}
D^{\dagger}(\xi)HD(\xi)D^{\dagger}(\xi)\Psi=ED^{\dagger}(\xi)\Psi.
\end{equation}
If we define the tilted Hamiltonian $H'=D^{\dagger}(\xi)HD(\xi)$ and the wave function $\Psi'=D^{\dagger}(\xi)\Psi$, this equation
can be written as
\begin{equation}
H'\Psi'=E\Psi'\label{h'}.
\end{equation}
With the displacement operator $D(\xi)$ and the Baker-Campbell-Hausdorff identity
\begin{equation}
e^{-A}Be^A=B+\frac{1}{1!}[B,A]+\frac{1}{2!}[[B,A],A]+\frac{1}{3!}[[[B,A],A],A]+...,
\end{equation}
we can find the similarity transformations
$D^{\dag}(\xi)K_{+}D(\xi)$, $D^{\dag}(\xi)K_{-}D(\xi)$ and
$D^{\dag}(\xi)K_{0}D(\xi)$ of the $su(1,1)$ Lie algebra generators.
These results are given by
\begin{equation}
D^{\dag}(\xi)K_{+}D(\xi)=\frac{\xi^{*}}{|\xi|}\alpha
K_{0}+\beta\left(K_{+}+\frac{\xi^{*}}{\xi}K_{-}\right)+K_{+},\label{simiK+}
\end{equation}
\begin{equation}
D^{\dag}(\xi)K_{-}D(\xi)=\frac{\xi}{|\xi|}\alpha
K_{0}+\beta\left(K_{-}+\frac{\xi}{\xi^{*}}K_{+}\right)+K_{-},\label{simiK-}
\end{equation}
\begin{equation}
D^{\dag}(\xi)K_{0}D(\xi)=(2\beta+1)
K_{0}+\frac{\alpha\xi}{2|\xi|}K_{+}+\frac{\alpha
\xi^*}{2|\xi|}K_{-},\label{simiK0}
\end{equation}
where $\alpha=\sinh(2|\xi|)$ and
$\beta=\frac{1}{2}\left[\cosh(2|\xi|)-1\right]$.
From these equations the tilted Hamiltonian becomes
\begin{eqnarray}
H'&=&K_0\left((2\beta+1)f+\frac{g\alpha\xi^*}{|\xi|}+\frac{g^*\alpha\xi}{|\xi|}\right)+
K_+\left(g(\beta+1)+\frac{f\alpha\xi}{2|\xi|}+\frac{g^*\beta\xi}{\xi^*}\right)+\nonumber\\
&&K_-\left(g^*(\beta+1)+\frac{f\alpha\xi^*}{2|\xi|}+\frac{g\beta\xi^*}{\xi}\right).
\end{eqnarray}
By choosing the coherent state parameters
\begin{equation}
\tau=\tanh^{-1}\left(2\gamma/f\right), \hspace{1 cm}\varphi=\phi,\label{parameters}
\end{equation}
the tilted Hamiltonian becomes proportional to $K_0$, since the coefficients of $K_{\pm}$ vanish. With these results we obtain
\begin{equation}
H'=\left(f^2-4\gamma^2\right)^{1/2}K_0.\label{htilt}
\end{equation}
Thus, since the wave function $\Psi'=D^{\dagger}(\xi)\Psi$ is an $SU(1,1)$ group state, by using equation (\ref{k0n}) the energy eigenvalues and the eigenstates of the original Hamiltonian $H$ are
\begin{equation}
E=(n+k)\left(f^2-4\gamma^2\right)^{1/2},\label{energy}
\end{equation}
\begin{equation}
|\zeta,k,n\rangle=D(\xi)|k,n\rangle.\label{DefPNCS}
\end{equation}
Notice that these equations are valid just in the case for which the relations (\ref{parameters}) are fulfilled.
For any complex value of the coherent state parameter $\zeta$, the states $|\zeta,k,n\rangle$ are what we call the $SU(1,1)$ Perelomov number coherent states. These states are a generalization of the standard Perelomov coherent states (\ref{PCN}), which are constructed from the ground state $|k,0\rangle$.

Due to the completeness of the group states $|k,n\rangle$, the $SU(1,1)$ Perelomov number coherent states satisfy the completeness relationship with respect to $n$
\begin{equation}
\sum_{n=0}^{\infty}|\zeta,k,n\rangle \langle\zeta,k,n|=I.
\end{equation}
Moreover, since the displacement operator $D(\xi)$ is unitary, the states $|\zeta,k,n\rangle$ are ortonormal since
\begin{equation}
\langle \zeta,k,n'|\zeta,k,n\rangle=\delta_{n',n}.
\end{equation}

Thus, the Perelomov number coherent states for an $SU(1,1)$ Hamiltonian (\ref{ham}) are constructed by applying the
displacement operator $D(\xi)$ to the excited group states (\ref{DefPNCS}). Furthermore, this definition has been used to study the Berry's phase for the degenerate parametric amplifier \cite{gerryberry} and for a weakly interacting Bose system \cite{huang}.

\subsection{$SU(1,1)$ Perelomov number coherent states for the discrete series}

Now, we shall construct the Perelomov number coherent states $|\zeta,k,n\rangle$ for any three generators $K_0$ and $K_\pm$ of the $su(1,1)$ Lie algebra. By using the normal form of the displacement operator $D(\xi)$ we obtain
\begin{equation}
|\zeta,k,n\rangle=D(\xi)|k,n\rangle=\exp(\zeta K_{+})\exp(\eta
K_{3})\exp(-\zeta^* K_{-})|k,n\rangle\label{defPCNS}.
\end{equation}
The action of each exponentials on an arbitrary excited group state $|k,n\rangle$ is obtained from equations (\ref{k+n})-(\ref{k0n}). We find they
are given by
\begin{equation}
\sum_{j=0}^\infty\frac{(-\zeta^* K_{-})^j}{j!}|k,n\rangle=
\sum_{j=0}^n\frac{(-\zeta^*)^j}{j!}\left(\frac{n!(2k+n-1)!}{(n-j)!(2k+n-1-j)!}\right)^{\frac{1}{2}}|k,n-j\rangle,
\end{equation}
\begin{equation}
\sum_{j=0}^\infty\frac{(\eta K_{0})^j}{j!}|k,n\rangle=\sum_{j=0}^\infty\frac{(\eta (k+n))^j}{j!}|k,n\rangle,
\end{equation}
\begin{equation}
\sum_{j=0}^\infty\frac{(\zeta K_{+})^j}{j!}|k,n\rangle=
\sum_{j=0}^\infty\frac{\zeta^j}{j!}\left(\frac{(n+j)!(2k+n+j-1)!}{n!(2k+n-1)!}\right)^{\frac{1}{2}}|k,n+j\rangle.
\end{equation}
From these results we obtain the $SU(1,1)$ Perelomov number coherent state
\begin{eqnarray}
|\zeta,k,n\rangle &=&\sum_{s=0}^\infty\frac{\zeta^s}{s!}\sum_{j=0}^n\frac{(-\zeta^*)^j}{j!}e^{\eta(k+n-j)}
\frac{\sqrt{\Gamma(2k+n)\Gamma(2k+n-j+s)}}{\Gamma(2k+n-j)}\nonumber\\
&&\times\frac{\sqrt{\Gamma(n+1)\Gamma(n-j+s+1)}}{\Gamma(n-j+1)}|k,n-j+s\rangle.\label{PNCS}
\end{eqnarray}
Notice that this equation does not depend on a particular setting of the parameters $\tau$ and $\varphi$. Equation (\ref{PNCS}) is not the only algebraic form for the Perelomov coherent number states, since Gerry gave an alternative representation of these coherent states. However, his coherent states are very difficult to work with, because they are given in terms of the confluent hypergeometric and Bargman V functions. To our knowledge the Gerry's results have not been explicitly calculated for any physical system. For example, if we set $n=0$ in equation (\ref{PNCS}), it is immediate to show that it reduces to standard Perelomov coherent states (\ref{PCN}). This result is more complicated to be obtained from the Gerry's expression (17) of reference \cite{gerryberry}.

Now, since $D(\xi)$ is unitary, the definition of the Perelomov number coherent states, equation (\ref{PNCS}) and equations (\ref{k+n})-(\ref{k0n}) lead to
\begin{equation}
L_{+}|\zeta,k,n\rangle=\sqrt{(n+1)(2k+n)}|\zeta,k,n+1\rangle,\label{l+n}
\end{equation}
\begin{equation}
L_{-}|\zeta,k,n\rangle=\sqrt{n(2k+n-1)}|\zeta,k,n-1\rangle,\label{l-n}
\end{equation}
\begin{equation}
L_{0}|\zeta,k,n\rangle=(k+n)|\zeta,k,n\rangle,\label{l0n}
\end{equation}
where we have defined the operators $L_{\pm}$ and $L_0$ as the similarity transformations
\begin{equation}
L_{\pm}=D(\xi)K_{\pm}D^{\dag}(\xi),\quad\quad L_{0}=D(\xi)K_{0}D^{\dag}(\xi)\label{antisimiL}.
\end{equation}
This implies that the operators $L_{\pm}$ and $L_{0}$
also satisfy the $su(1,1)$ commutation relations
\begin{eqnarray}
[L_{0},L_{\pm}]=\pm L_{\pm},\quad\quad [L_{-},L_{+}]=2L_{0}.\label{comL}
\end{eqnarray}
Results above are very interesting since they show that $L_{\pm}$
and $L_{0}$ contain the same algebraic structure that operators
$K_{\pm}$ and $K_{0}$, being the states $|\zeta,k,n\rangle$ the new basis to get the
unitary irreducible representation for the $su(1,1)$ Lie algebra.

Since $D^{\dag}(\xi)=D(-\xi)$, the explicit form of the operators
$L_{\pm}$ and $L_0$ are those obtained from the similarity transformations
(\ref{simiK+}), (\ref{simiK-}) and (\ref{simiK0})
\begin{equation}
L_{+}=-\frac{\xi^{*}}{|\xi|}\alpha
K_{0}+\beta\left(K_{+}+\frac{\xi^{*}}{\xi}K_{-}\right)+K_{+},
\end{equation}
\begin{equation}
L_{-}=-\frac{\xi}{|\xi|}\alpha
K_{0}+\beta\left(K_{-}+\frac{\xi}{\xi^{*}}K_{+}\right)+K_{-},
\end{equation}
\begin{equation}
L_{0}=(2\beta+1)K_{0}-\frac{\alpha\xi}{2|\xi|}K_{+}-\frac{\alpha \xi^*}{2|\xi|}K_{-}.
\end{equation}

The time evolution operator $U(t)$ for an arbitrary Hamiltonian $H$
is defined as $U(t)=e^{-iHt/\hbar}$ \cite{Cohen}. Therefore, the time evolution of the
$SU(1,1)$ Perelomov number coherent states is given by
\begin{equation}
|\zeta(t),k,n\rangle=e^{-iHt/\hbar}|\zeta,k,n\rangle=D(\xi)D^{\dagger}(\xi)e^{-iHt/\hbar}D(\xi)|k,n\rangle.
\end{equation}
If we consider a Hamiltonian of the form given by equation (\ref{ham}), we obtain
\begin{equation}
D^{\dagger}(\xi)e^{-iHt/\hbar}D(\xi)=e^{-iH't/\hbar}.
\end{equation}
Thus, if we define $\Omega=\left(f^2-4\gamma^2\right)^{1/2}$, from equations (\ref{k0n}) and (\ref{htilt})
\begin{equation}
|\zeta(t),k,n\rangle=e^{-i\Omega(k+n)t/\hbar}|\zeta, k,n\rangle.
\end{equation}
Notice that in order to obtain this result, we have restricted the coherent state parameters to those given in equation (\ref{parameters}). Therefore, the time evolution of the $SU(1,1)$ generalized coherent states is obtained by adding the phase $e^{-i\Omega(k+n)t/\hbar}$ to the stationary Perelomov number coherent states (\ref{DefPNCS}).

The time evolution operator of the non-degenerate parametric amplifier was previously studied by Rekdal and Skagerstam \cite{rekdal}.
They obtained the most general form of the time evolution operator for this problem, which is expressed by means of the three $su(1,1)$ lie algebra generators. Due to the tilting transformation, our time evolution operator is written in the shortest way in terms of just the third $su(1,1)$ lie algebra generator $K_0$.

\section{The nondegenerate parametric amplifier.}

As a particular case of our motivating example, we consider the Hamiltonian which describes the stationary nondegenerate parametric amplifier in quantum optics \cite{mandel,walls,scully}
\begin{equation}
H=\omega\left(a^{\dag}a+b^{\dag}b\right)+\chi\left[a^{\dag}b^{\dag}e^{-i\Phi}-abe^{i\Phi}\right],
\end{equation}
where the coupling constant $\chi$ is proportional to the second-order susceptibility of the medium and to the amplitude of the pump,
$\Phi$ is the phase of the pump field and $\omega$ is its frequency.
For the $SU(1,1)$ two-mode harmonic oscillator realization the Lie algebra may be spanned as \cite{gerrytwomode}
\begin{eqnarray}
K_0=\frac{1}{2}\left(a^{\dag}a+b^{\dag}b+1\right),\quad K_+=a^{\dag}b^{\dag},\quad K_-=ab,\label{alreal}
\end{eqnarray}
where $a$ and $b$ are the annihilation operators of a right and left circular quantum respectively. With this realization we can write the nondegenerate parametric amplifier
Hamiltonian as
\begin{equation}
H=2\omega K_0+\chi K_+e^{-i\Phi}-\chi K_-e^{i\Phi}-\omega.
\end{equation}
Therefore, by choosing $\tau=\tanh^{-1}\left(2\chi/\omega\right)$ and $\phi=\Phi$ we obtain the tilted Hamiltonian $H'$
\begin{equation}
H'=2(\omega^2-\chi^2)^{1/2}K_0-\omega=\frac{(\omega^2-\chi^2)^{1/2}}{\omega}H_{HO}-\omega,\label{energyparametric}
\end{equation}
where $H_{HO}$ is the Hamiltonian of the two-dimensional harmonic oscillator. Thus, the eigenfunctions of the Hamiltonian $H'$ are those of the
two-dimensional harmonic oscillator \cite{schwinger}
\begin{equation}
\psi'_{N,m}(r,\phi)=\frac{1}{\sqrt{\pi}}e^{im\phi}(-1)^{\frac{N-m}{2}}\sqrt{\frac{2\left(\frac{N-m}{2}\right)!}{\left(\frac{N+m}{2}\right)!}}
r^{m}L_{\frac{1}{2}(N-m)}^{m}(r^2)e^{-\frac{1}{2}r^2},
\end{equation}
or
\begin{equation}
\psi'_{n_{r},m}(r,\phi)=\frac{1}{\sqrt{\pi}}e^{im\phi}(-1)^{n_{r}}\sqrt{\frac{2\left(n_{r}\right)!}{\left(n_{r}+m\right)!}}
r^{m}L_{n_{r}}^{m}(r^2)e^{-\frac{1}{2}r^2},\label{function}
\end{equation}
where $n_{r}=\frac{1}{2}(N-m)$ is the radial quantum number. In polar coordinates the creation and annihilation operators take the form
\begin{eqnarray}
a=\frac{e^{-i\varphi}}{2}\left[r+\frac{\partial}{\partial r}-\frac{i}{r}\frac{\partial}{\partial \varphi}\right],\quad\quad
a^{\dag}=\frac{e^{i\varphi}}{2}\left[r-\frac{\partial}{\partial r}-\frac{i}{r}\frac{\partial}{\partial \varphi}\right],\label{apolar}
\end{eqnarray}
\begin{eqnarray}
b=\frac{e^{i\varphi}}{2}\left[r+\frac{\partial}{\partial r}+\frac{i}{r}\frac{\partial}{\partial \varphi}\right],\quad\quad
b^{\dag}=\frac{e^{-i\varphi}}{2}\left[r-\frac{\partial}{\partial r}+\frac{i}{r}\frac{\partial}{\partial \varphi}\right].\label{bpolar}
\end{eqnarray}
The action of these operators on the basis $|N,m\rangle$ is given by \cite{wallace}
\begin{eqnarray}
a|N,m\rangle=\sqrt{\frac{1}{2}(N+m)}|N-1,m-1\rangle,\quad a^{\dag}|N,m\rangle=\sqrt{\frac{1}{2}(N+m)+1}|N+1,m+1\rangle,\label{acta}
\end{eqnarray}
\begin{eqnarray}
b|N,m\rangle=\sqrt{\frac{1}{2}(N-m)}|N-1,m+1\rangle,\quad b^{\dag}|N,m\rangle=\sqrt{\frac{1}{2}(N-m)+1}|N+1,m-1\rangle.\label{actb}
\end{eqnarray}
Thus, by substituting equations (\ref{apolar}) and (\ref{bpolar}) into equation (\ref{alreal}), we obtain the $su(1,1)$ Lie algebra operators
\begin{equation}
K_+=\frac{1}{4}\left[r^2-2r\frac{\partial}{\partial r}-2+\frac{\partial^2}{\partial r^2}+\frac{1}{r}\frac{\partial}{\partial r}+\frac{1}{r^2}\frac{\partial^2}{\partial \varphi^2}\right],
\end{equation}
\begin{equation}
K_-=\frac{1}{4}\left[r^2+2r\frac{\partial}{\partial r}+2+\frac{\partial^2}{\partial r^2}+\frac{1}{r}\frac{\partial}{\partial r}+\frac{1}{r^2}\frac{\partial^2}{\partial \varphi^2}\right],
\end{equation}
\begin{equation}
K_0=\frac{1}{4}\left[r^2-\frac{\partial^2}{\partial r^2}-\frac{1}{r}\frac{\partial}{\partial r}+\frac{1}{r^2}\frac{\partial^2}{\partial \varphi^2}\right].
\end{equation}
The Casimir operator $K^2$ for this algebra realization is
\begin{equation}
K^2=-\frac{1}{4}\left(1+\frac{\partial^2}{\partial \varphi^2}\right).
\end{equation}
The application of this operator to any eigenfunction of the two-dimensional isotropic harmonic oscillator leads to
\begin{equation}
C\psi=\frac{1}{4}(m^2-1)\psi=k(k-1)\psi.
\end{equation}
Thus, the Bargmann index is $k=\frac{1}{2}(m+1)$. The other group index $n$, is obtained by noting that
\begin{equation}
K_0|N,m\rangle=\frac{1}{2}\left(a^{\dag}a+b^{\dag}b+1\right)|N,m\rangle=\frac{1}{2}(N+1)|N,m\rangle,\label{actk0}
\end{equation}
where we have used equations (\ref{acta}) and (\ref{actb}). By comparison of equations (\ref{k0n}) and (\ref{actk0}) we deduce $n=\frac{1}{2}(N-m)$, which
is equals to the radial quantum number $n_{r}$.

The Perelomov number coherent states for the two-dimensional harmonic oscillator, and therefore, the eigenfunctions of the nondegenerate parametric amplifier are obtained by substituting the states (\ref{function}) into equation (\ref{PNCS}). With the definition $l=m+1/2$, we obtain
\begin{eqnarray}
\langle r,\varphi|\zeta,k,n\rangle &=&\frac{(-1)^{n}}{\sqrt{\pi}}e^{i(l-1/2)\phi}\sum_{s=0}^\infty\frac{\zeta^s}{s!}\sum_{j=0}^n\frac{(-\zeta^*)^j}{j!}e^{\eta(k+n-j)}
\frac{\sqrt{2\Gamma(n+1)\Gamma(n+l+\frac{1}{2})}}{\Gamma(n-j+l+\frac{1}{2})}\nonumber\\
&&\times\frac{\Gamma(n-j+s+1)}{\Gamma(n-j+1)}e^{-r^2/2}r^{(l-1/2)}L_{n-j+s}^{l-1/2}(r^2).
\end{eqnarray}
By interchanging the order of summations and using the sum $48.7.6$ of \cite{hansen}, we find
\begin{eqnarray}
\langle r,\varphi|\zeta,k,n\rangle &=&\frac{(-1)^{n}}{\sqrt{\pi}}e^{i(l-1/2)\phi}\frac{(1-|\zeta|^2)^{n+\frac{l}{2}+\frac{1}{4}}}{(1-\zeta)^{n+l+\frac{1}{2}}}e^{-\frac{r^2(\zeta+1)}{2(1-\zeta)}}
r^{l-1/2}\nonumber\\
&&\times\sum_{j=0}^n\frac{\sqrt{2\Gamma(n+1)\Gamma(n+l+\frac{1}{2})}}{\Gamma(j+1)\Gamma(n-j+l+\frac{1}{2})}
\left(\frac{(1-\zeta)(-\zeta^*)}{1-|\zeta|^2}\right)^jL_{n-j}^{l-1/2}\left(\frac{r^2}{1-\zeta}\right),
\end{eqnarray}
where we have used that $\eta=-2\ln \cosh|\xi|=\ln(1-|\zeta|^2)$. The remaining sum can be calculated by using the formula $48.7.8$ of \cite{hansen}.
This allows to obtain the eigenfunctions of the non-degenerate parametric amplifier
\begin{eqnarray}
\psi_{n,m}&=&\sqrt{\frac{2\Gamma(n+1)}{\Gamma(n+m+1)}}\frac{(-1)^n}{\sqrt{\pi}}e^{im\phi}
\frac{(-\zeta^*)^n(1-|\zeta|^2)^{\frac{m}{2}+\frac{1}{2}}(1+\sigma)^n}{(1-\zeta)^{m+1}}\nonumber\\
&&\times e^{-\frac{r^2(\zeta+1)}{2(1-\zeta)}}r^{m}L_n^{m}\left(\frac{r^2\sigma}{(1-\zeta)(1-\sigma)}\right),
\end{eqnarray}
where we have used $m=l-1/2$ and defined
\begin{equation}
\sigma=\frac{1-|\zeta|^2}{(1-\zeta)(-\zeta^*)}.
\end{equation}
Moreover, the time evolution of these states is obtained by adding the phase $e^{-i\Omega\left(n+\frac{m}{2}+\frac{1}{2}\right)t/\hbar}$ to them.
When $n=0$, these states are reduced to the standard Perelomov coherent states for the two-dimensional harmonic oscillator
\begin{eqnarray}
\psi_{0}=\sqrt{\frac{2}{\pi\Gamma(m+1)}}
\frac{(1-|\zeta|^2)^{\frac{m}{2}+\frac{1}{2}}}{(1-\zeta)^{m+1}}e^{-\frac{r^2(\zeta+1)}{2(1-\zeta)}}e^{im\phi}r^{m},
\end{eqnarray}
which are the ground states for the non-degenerate parametric amplifier. The radial part of these coherent states is in full agreement
with those previously reported for the Landau levels \cite{apolonica}.

The energy spectrum of this system can be easily computed by using equation (\ref{energyparametric})
\begin{equation}
E=2(\omega^2-\chi^2)^{1/2}\left(n+\frac{m}{2}+\frac{1}{2}\right)-\omega.
\end{equation}
If we set $\chi=0$, this energy spectrum reduces to that of the two-dimensional harmonic oscillator.
Thus, the tilting transformation and our Perelomov number coherent states (\ref{PNCS}) allowed us to calculate the eigenfunctions and the energy spectrum
of the nondegenerate parametric amplifier. To our knowledge these results have not been published.

\section{Concluding remarks}

We obtained an algebraic form of the $SU(1,1)$ Perelomov number coherent states for the discrete series. We construct
a set of operators which close the $su(1,1)$ Lie algebra, being the generalized coherent states the basis for
its unitary irreducible representation.

By using the tilting transformation, we showed that the most general $SU(1,1)$ coherence-preserving Hamiltonian
has the Perelomov number coherent states as eigenfunctions. We applied our results to obtain the energy spectrum and
the eigenfunctions of the non-degenerate parametric amplifier, which are shown to be the Perelomov number coherent states of the
two-dimensional harmonic oscillator.

An alternative representation for the Perelomov number coherent states developed in this work was previously
obtained by Gerry in terms of the hypergeometric and Bargmann V functions. However,
our representation allowed us to calculate these coherent states for the two-dimensional harmonic
oscillator explicitly. Finally, our formulation of the Perelomov number coherent states can be applied to Hamiltonians of other
quantum systems and will be reported in a future work.

\section*{Acknowledgments}
This work was partially supported by SNI-M\'exico, COFAA-IPN,
EDI-IPN, EDD-IPN, SIP-IPN project number $20130620$.

\end{document}